\documentclass[aps,10pt,prd,notitlepage,showpacs,nofootinbib,superscriptaddress, compressed]{revtex4-1}

\pdfoutput=1 
\usepackage[utf8]{inputenc}
\usepackage[english]{babel}
\usepackage{amsmath}
\usepackage{graphicx}
\usepackage{dcolumn}
\usepackage{pbox}
\usepackage{amssymb}
\usepackage{epsfig}
\usepackage[dvipsnames]{xcolor}
\usepackage{slashed}
\usepackage{amssymb}
\usepackage{mathrsfs}
\usepackage{rotating}
\usepackage{color}
\usepackage[normalem]{ulem}
\usepackage{multirow}
\usepackage[font=small]{caption}
\usepackage[font=small]{subcaption}
\usepackage{url}
\usepackage{braket}
\definecolor{DarkBlue}{rgb}{0.7, 0.4, 1} 
\definecolor{Blue}{rgb}{0, 0.8, 0} 
\definecolor{MyLightBlue}{rgb}{0.5,0.7,1.9}
\definecolor{MyGreen}{rgb}{0.0,0.2, 0.0}
\definecolor{MyBrickRed}{rgb}{0, 0.5, 0.2}
\RequirePackage{hyperref}
\hypersetup{colorlinks, citecolor=Blue,linkcolor=DarkBlue, urlcolor=Green}

\newcommand{\bea}{\begin{eqnarray}}
\newcommand{\eea}{\end{eqnarray}}

\makeatletter

\makeatletter
\renewcommand\@makecaption[2]{%
  \par
  
  \vskip\abovecaptionskip
  \begingroup
  
   \small\rmfamily
    \begingroup
     \samepage
     \flushing
     \let\footnote\@footnotemark@gobble
     \@make@capt@title{#1}{#2}\par
    \endgroup
  \endgroup
  \vskip\belowcaptionskip
}
\makeatother

\DeclareUnicodeCharacter{2212}{-}
\begin{document}
\title{Resonant leptogenesis in minimal $U(1)_X$ extensions of the Standard Model}
\author{Arindam Das}
\email{adas@particle.sci.hokudai.ac.jp}
\affiliation{Institute for the Advancement of Higher Education, Hokkaido University, Sapporo 060-0817, Japan}
\affiliation{Department of Physics, Hokkaido University, Sapporo 060-0810, Japan}
\author{Yuta Orikasa}\email{yuta.orikasa@utef.cvut.cz}
\affiliation{Institute of Experimental and Applied Physics, Czech Technical University in Prague, Husova 240/5, 110 00 Prague 1, Czech Republic}
\begin{abstract}
We investigate a general $U(1)_X$ scenario where we introduce three generations of Standard Model (SM) singlet Right Handed Neutrinos (RHNs) to generate the light neutrino mass through the seesaw mechanism after the breaking of $U(1)_X$ and electroweak symmetries. In addition to that, a general $U(1)_X$ scenario involves an SM-singlet scalar field and due to the $U(1)_X$ symmetry breaking the mass of a neutral beyond the SM (BSM) gauge boson $Z^\prime$ is evolved. The RHNs, charged under the $U(1)_X$ scenario, can explain the origin of observed baryon asymmetry through the resonant leptogenesis process. Applying observed neutrino oscillation data we study $Z^\prime$ and BSM scalar-induced processes to reproduce the observed baryon asymmetry. Hence we estimate bounds on the $U(1)_X$ gauge coupling $(g_X)$ and the mass of the $Z^\prime$ $(M_{Z^\prime})$ for different $U(1)_X$ charges and benchmark masses of the RHN and SM-singlet scalar. Finally, we compare our results with limits obtained from the existing limits from LEP-II and LHC. We find that depending on the $U(1)_X$ charges, the masses of RHNs and SM-singlet scalar resonant leptogenesis could provide a stronger limit on $g_X$ for $M_{Z^\prime} > 5.8$ TeV which could be probed by high energy scattering experiment in future.
\noindent 
\end{abstract}
\maketitle
\section{Introduction}
The origin of baryon asymmetry of the universe is an unsolved puzzle of modern cosmology. The WIMP satellite \cite{WMAP:2010qai} observed that the ratio of the baryon minus anti-baryon density $(n_{B-\bar{B}})$ over the entropy density of the universe $(s)$ is $Y_B= 8.7\times 10^{-11}$ at an accuracy of $10\%$ precision level. A strong first-order phase transition could be required to explain the origin of baryon (B) asymmetry induced by the electroweak baryogenesis \cite{Cohen:1993nk,Funakubo:1996dw,Trodden:1998ym} within the Standard Model (SM) framework, however, the observation of the SM Higgs mass around 125 GeV \cite{CMS:2012qbp,ATLAS:2012yve} does not indicate such a phenomenon. As a result, electroweak baryogenesis is possibly ruled out in the SM scenario. 

In addition to the anomaly associated with the observed baryon asymmetry there are interesting developments in the lepton sector of the SM which are observed in the form of tiny neutrino mass, neutrino flavor oscillation and mixing \cite{ParticleDataGroup:2022pth} and can not be explained within the SM framework. As a result an approach of stepping beyond the SM (BSM) was important which evolved in the form of seesaw mechanism \cite{Minkowski:1977sc,Yanagida:1979as,Gell-Mann:1979vob,Mohapatra:1979ia,Glashow:1979nm} where the origin of tiny neutrino mass could be explained through the suppression of a heavy Majorana mass scale. In this case the SM is extended by SM-singlet heavy Majorana type Right Handed Neutrinos (RHNs) at the intermediate scale which create lepton number violation by unit 2 through a dimension five Weinberg operator \cite{Weinberg:1979sa}.

An attempt to connect the baryon asymmetry of the universe and possibility of tiny neutrino mass could be established by the seesaw scenario by an attractive possibility called leptogenesis \cite{Fukugita:1986hr}. The out of equilibrium decay of the Majorana type RHNs can generate lepton (L) asymmetry in the universe which can be converted into baryon asymmetry through the sphaleron transition violating $(B+L)$ quantum number \cite{Manton:1983nd,Klinkhamer:1984di}. Hence, the ratio between the baryon and lepton conversion rate can be $-\frac{28}{79}$ following \cite{Khlebnikov:1988sr,Iso:2010mv,Chauhan:2021xus} for three (one) generations of SM fermions (Higgs). In this context, we mention that thermal leptogenesis imparts a lower bound on the Majorana neutrinos $(M_N)$ to generate sufficient amount of baryon asymmetry as $M_N > 10^9$ GeV \cite{Plumacher:1996kc,Buchmuller:2002rq,Buchmuller:2004nz}. Such heavy Majorana neutrinos can not be observed at the hadron and lepton colliders. 

In order to test heavy Majorana neutrinos at high energy colliders we need to think possible alternative scenarios which may explain the origin of small neutrino mass and observed baryon asymmetry simultaneously. General U$(1)$ extensions of the SM are such scenarios in which these aspects can be addressed nicely. In this case, the SM is extended by a general $U(1)_X$ gauge group where three generations of the RHNs are introduced with an SM-singlet BSM scalar. These RHNs not only help to generate the tiny neutrino mass though the seesaw mechanism after the general $U(1)$ symmetry breaking, but also cancel the gauge gravity anomalies \cite{Oda:2015gna,Das:2016zue}. If general $U(1)$ symmetry breaking scale lies around the TeV scale, the corresponding neutral gauge boson commonly known as $Z^\prime$ and the RHNs production at high energy colliders. We extensively studied the RHN production form $Z^\prime$ at different hadron and lepton colliders in \cite{Das:2022rbl} applying detector simulations which are not taken into account in a recent article \cite{Chauhan:2024jfq} while following the flow-chart of \cite{Das:2022rbl} with inadequate insights on the results. In this scenario there is another interesting aspect which affects the $Z^\prime$ searches. After the anomaly cancellation we find that the left and right handed fermions of the model interact differently with the $Z^\prime$ boson. The general $U(1)$ scenarios considered in this article are at the TeV scale prohibiting thermal leptogenesis to occur, because in this case the RHNs are lighter than the bound of $10^{10}$ GeV. The CP asymmetry parameter in this case is proportional to the modulus squared of the Dirac Yukawa coupling $(Y_D)$ between the RHNs and the SM lepton doublet. Due to the smallness of $Y_D$, the right amount of baryon asymmetry of the universe can not be generated. It has also been found that if two RHNs are almost degenerate in masses, an enhancement \cite{Flanz:1996fb} of the CP asymmetry parameter can take place making the leptogenesis scenario viable for the TeV scale RHNs. This is called resonant leptogenesis \cite{Pilaftsis:1997jf,Pilaftsis:2003gt}. In this case maximum attainable enhancement could be achieved if mass difference between the two generations of the RHNs is $(\mathcal{O}(\Gamma_{N_i}))$, the total decay width $(\Gamma_{N_i})$ of either of the generations of the RHNs $(N_i)$. Hence in principle by tuning the mass difference between any two generations of the RHNs, the CP asymmetry could be attained around $\mathcal{O}(1)$, however, in the presence of general $U(1)$ extensions the lepton asymmetry through RHN decay is suppressed due to the interaction with the TeV scale $Z^\prime$ gauge boson. As a result such a scenario makes the generation of the CP asymmetry of the universe non-trivial \cite{Blanchet:2009bu}. 

In this paper, we study resonant leptogenesis in general $U(1)$ extensions of the SM where $U(1)_X$ and $U(1)_{q+xu}$ symmetry breaks at the TeV scale. Solving Boltzmann equations and considering different model parameters we show the dependence of baryon asymmetry via leptogenesis on neutrino Dirac Yukawa couplings and TeV scale nearly degenerate heavy Majorana neutrino masses. In this analysis we involve neutrino oscillation parameters in order to relate the baryon asymmetry via leptogenesis with realistic data. In this analysis we also study the role of the Dirac CP phase in the $U_{\rm{PMNS}}$ matrix in order to explain baryon asymmetry of the universe. 

We organize the paper in the following way. In Sec.~\ref{model} we describe the model. In Sec.~\ref{secIII} we analyse the resonant leptogenesis scenario. In Sec.~\ref{secIV} we apply neutrino oscillation data to investigate the role of Dirac CP phase to generate the right amount of baryon asymmetry of the universe. We discuss our results in Sec.~\ref{secIV} and conclude it in Sec.~\ref{secV}.
\section{General $U(1)$ extensions of Standard Model}
\label{model}
General $U(1)$ extensions of the SM involve an SM-singlet scalar $\Phi$ and three generations of the SM-singlet RHNs. Apart from participating in the neutrino mass generation mechanism, the RHNs also help to cancel gauge and mixed gauge-gravity anomalies. The particle content is given in Tab.~\ref{tab1} where generation independent $U(1)_X$ charges $(x_f, f= \{ q,~u,~d,~\ell,~e,~\nu \})$ of the particles are given. 
\begin{table}[t]
\begin{center}
\begin{tabular}{||c|ccc||c||c||}
\hline
\hline
            & SU(3)$_c$ & SU(2)$_L$ & U(1)$_Y$ & U(1)$_X$  & U(1)$_{q+xu}$ \\[2pt]
            \hline
\hline
&&&&\\[-12pt]
$q_L^i$    & {\bf 3}   & {\bf 2}& $\frac{1}{6}$ & 		 $x_q=\frac{1}{6}x_H + \frac{1}{3}x_\Phi$    &   $\frac{1}{3}$   \\[2pt]
$u_R^i$    & {\bf 3} & {\bf 1}& $\frac{2}{3}$ &  	  $x_u=\frac{2}{3}x_H + \frac{1}{3}x_\Phi$  & 		 $\frac{x}{3}$   \\[2pt]
$d_R^i$    & {\bf 3} & {\bf 1}& $-\frac{1}{3}$ & 	 $x_d=-\frac{1}{3}x_H + \frac{1}{3}x_\Phi$  & 		 $\frac{2-x}{3}$   \\[2pt]
\hline
\hline
&&&&\\[-12pt]
$\ell_L^i$    & {\bf 1} & {\bf 2}& $-\frac{1}{2}$ & 	 $x_\ell=- \frac{1}{2}x_H - x_\Phi$   & 		 $-1$   \\[2pt]
$e_R^i$   & {\bf 1} & {\bf 1}& $-1$   &		$x_e=- x_H - x_\Phi$   & 		 $-(\frac{2+x}{3})$   \\[2pt]
\hline
\hline
$N_R$   & {\bf 1} & {\bf 1}& $0$   &	 $x_\nu=- x_\Phi$  & 		 $\frac{-4+x}{3}$   \\[2pt]
\hline
\hline
&&&&\\[-12pt]
$H$         & {\bf 1} & {\bf 2}& $-\frac{1}{2}$  &  	 $\frac{x_H}{2}$&  	 $\frac{1-x}{3}$\\ [2pt]
$\Phi$      & {\bf 1} & {\bf 1}& $0$  & 	 $2 x_\Phi$ & 	 $2 (\frac{-4+x}{3})$ \\ [2pt]
\hline
\hline
\end{tabular}
\end{center}
\caption{
Particle content of the general $U(1)$ extensions of the SM where $i(=1, 2, 3)$ represents the family index. The quantities $x_H$, $x_\Phi$ and $x$ are the real parameters.}
\label{tab1}
\end{table}
We write the Yukawa interactions following $\mathcal{G}_{\rm SM} \otimes$ $U(1)_X$ gauge symmetry as
\begin{equation}
{\cal L}^{\rm Yukawa} = - Y_u^{\alpha \beta} \overline{q_L^\alpha} H u_R^\beta
                                - Y_d^{\alpha \beta} \overline{q_L^\alpha} \tilde{H} d_R^\beta
				 - Y_e^{\alpha \beta} \overline{\ell_L^\alpha} \tilde{H} e_R^\beta
				- Y_D^{\alpha \beta} \overline{\ell_L^\alpha} H N_R^\beta- Y_N^\alpha \Phi \overline{(N_R^\alpha)^c} N_R^\alpha + {\rm H.c.}~,
\label{LYk}   
\end{equation}
where $\alpha, \beta$ are the generation indices indicating three generations of the fermions involved in the theory. Here $H$ is the SM Higgs doublet, and $\tilde{H} = i  \tau^2 H^*$ with $\tau^2$ being the second Pauli matrix. Hence using $U(1)_X$ neutrality we can write the following conditions as 
\begin{eqnarray}
-\frac{1}{2} x_H^{} &=& - x_q + x_u \ =\  x_q - x_d \ =\  x_\ell - x_e=\ - x_\ell + x_\nu~;~
2 x_\Phi^{}	= - 2 x_\nu~. 
\label{Yuk}
\end{eqnarray} 
Using the gauge and mixed gauge-gravity anomaly cancellation conditions we can relate general $U(1)_X$ charges of the fermions in the following way 
\begin{align}
{\rm U}(1)_X \otimes \left[ {\rm SU}(3)_C \right]^2&\ :&
			2x_q - x_u - x_d &\ =\  0, \nonumber \\
{\rm U}(1)_X \otimes \left[ {\rm SU}(2)_L \right]^2&\ :&
			3x_q + x_\ell &\ =\  0, \nonumber \\
{\rm U}(1)_X \otimes \left[ {\rm U}(1)_Y \right]^2&\ :&
			x_q - 8 x_u - 2 x_d + 3x_\ell - 6 x_e &\ =\  0, \nonumber \\
\left[ {\rm U}(1)_X \right]^2 \otimes {\rm U}(1)_Y &\ :&
			{x_q}^2 - {2x_u}^2 + {x_d}^2 - {x_\ell}^2 + {x_e}^2 &\ =\  0, \nonumber \\
\left[ {\rm U}(1)_X \right]^3&\ :&
			{6x_q}^3 - {3x_u}^3 - {3x_d}^3 + {2x_\ell}^3 - {x_\nu}^3 - {x_e}^3 &\ =\  0, \nonumber \\
{\rm U}(1)_X \otimes \left[ {\rm grav.} \right]^2&\ :&
			6x_q - 3x_u - 3x_d + 2x_\ell - x_\nu - x_e &\ =\  0, 
\label{anom-f}
\end{align}
respectively. Solving Eqs.~\eqref{Yuk} and \eqref{anom-f} we express the fermionic $U(1)_X$ charges in terms of $x_H$ and $x_\Phi$ and the $U(1)_X$ charges of the SM fermions can be written as linear combination of the $U(1)_Y$ and B$-$L charges. Hence we find that left and right handed fermions are differently charged under the $U(1)_X$ gauge group and as a result they interact differently with the associated neutral BSM gauge boson $Z^\prime$ manifesting a chiral nature of the model. Such a chiral behavior affect any interaction process mediated by the $Z^\prime$ associating the fermions of this model. In this scenario, if we fix $x_\Phi^{}=1$ and $x_H^{}=0$ we can reproduce the B$-$L scenario and with $x_H^{}=-2$ we find that the $U(1)_X$ charges of the left handed fermions become zero, however, the right handed fermions have nonzero $U(1)_X$ charges. This is called $U(1)_{\rm R}$ scenario. In addition to that there are other properties of this model if we change $x_H$ keeping $x_\Phi=1$ fixed as 1. For example, in case of $x_H^{}=-1$, the $U(1)_X$ charge of $e_R$ vanishes, in case of $x_H^{}=-0.5$, the U$(1)_X$ charge of $u_R$ vanishes and in case of  $x_H^{}=1$, the $U(1)_X$ charge of $d_R^{}$ vanishes, respectively. However, for these respective $x_H$ values remaining fermions interact with the $Z^\prime$. The $U(1)_X$ gauge coupling is considered as $g_X$ for this model. This is a free parameter and it appears as either $g^\prime x_H^{}$ or $g^\prime x_\Phi^{}$ in the kinetic part of the Lagrangian. The other scenario $U(1)_{q+xu}$ reduces to B$-$L when $x=1$ and it can not reduce to the $U(1)_R$ case because $U(1)_{q+xu}$ charges of the SM right handed fermions do not become zero simultaneously.  

The renormalizable scalar potential in this scenario can be written as
\begin{align}
  V \ = \ m_H^2(H^\dag H) + \lambda_H^{} (H^\dag H)^2 + m_\Phi^2 (\Phi^\dag \Phi) + \lambda_\Phi^{} (\Phi^\dag \Phi)^2 + \lambda_{\rm mix} (H^\dag H)(\Phi^\dag \Phi)~,
\end{align}
where the fields $\Phi$ and $H$ are SM-singlet scalar and SM Higgs doublet, respectively. These scalar fields can be approximated separately in the analysis of scalar potential where $\lambda_{\rm mix}$ is very small \cite{Oda:2015gna,Das:2016zue}. After the breaking of $U(1)_X^{}$ gauge symmetry and electroweak symmetry, the scalar vacuum expectation values (VEVs) for the fields $H$ and $\Phi$ are developed as 
\begin{equation}
  <H> \ = \ \frac{1}{\sqrt{2}}\begin{pmatrix} v+h\\0 
  \end{pmatrix}, \quad {\rm and}\quad 
  <\Phi> \ =\  \frac{v_\Phi^{}+\phi}{\sqrt{2}},
\end{equation}
where $v=246$ GeV is marked as electroweak scale at the potential minimum and $v_\Phi^{}$ is considered as a free parameter. After the $U(1)_X$ gauge symmetry is broken with a limit $v_\Phi^{} \gg v$, the mass of $Z^\prime$ is evolved and can be defined as 
\begin{equation}
 M_{Z^\prime}^{}=  2 g_X^{}  v_\Phi^{} x_\Phi
\end{equation}
which is a free parameter in this model. From Eq.~\eqref{LYk}, we find that the RHNs interact with $\Phi$ which can generate the Majorana mass for heavy neutrinos after $U(1)_X$ symmetry is broken. After the electroweak symmetry breaking, the Dirac mass term is generated from the Yukawa coupling involving left handed lepton and SM Higgs doublets respectively. These mass terms finally allow seesaw mechanism to explain the origin of tiny neutrino mass and flavor mixing. We write the Majorana and Dirac mass terms from Eq.~\ref{LYk} as
\begin{equation}
    m_{N^\alpha}^{} \ = \ \frac{Y^\alpha_{N}}{\sqrt{2}} v_\Phi^{}, \, \, \, \, \,
    m_{D}^{\alpha \beta} \  =  \ \frac{Y_{D}^{\alpha \beta}}{\sqrt{2}} v,
\label{mDI}
\end{equation}
respectively. 

\subsection{$Z^\prime$ interactions with the fermions}
General $U(1)$ extensions of the SM allow a neutral BSM gauge boson which can interact with the SM fermions $(f)$. The interaction Lagrangian involves $U(1)_X$ charges manifesting chiral behavior as 
\begin{eqnarray}
\mathcal{L}^{\rm{int}} = -g_X (\overline{f}\gamma_\mu q_{f_{L}^{}}^{} P_L^{} f+ \overline{f}\gamma_\mu q_{f_{R}^{}}^{}  P_R^{} f) Z_\mu^\prime~,
\label{Lag1}
\end{eqnarray}
where $P_{L(R)}^{}= (1 \pm \gamma_5)/2$  is the projection operator, $q_{L(R)}$ is the corresponding general $U(1)$ charge of the left (right) handed fermions which could be found from Tab.~\ref{tab1}. These quantities depend on $x_H$ and $x_\Phi$ respectively. Hence we calculate partial decay widths of $Z^\prime$ into charged fermions as 
\begin{align}
\label{eq:width-ll}
    \Gamma(Z' \to \bar{f} f)
    &= N_C^{} \frac{ g_{X}^2 M_{Z'}^{}}{24 \pi} \left[ \left( q_{f_L^{}}^2 + q_{f_R^{}}^2 \right) \left( 1 - \frac{m_f^2}{M_{Z'}^2} \right) + 6 q_{f_L^{}}^{} q_{f_R^{}}^{} \frac{m_f^2}{M_{Z'}^2} \right] \sqrt{1-4\frac{m_f^2}{M_{Z^\prime}^2}}~,
\end{align}    
where $m_f$ is the SM fermion mass and $N_C^{}=1~(3)$ for the SM leptons (quarks). The partial decay width of $Z^\prime$ into a pair of light neutrinos for three generations neglecting neutrino masses is given by
\begin{align}   
\label{eq:width-nunu}
    \Gamma(Z' \to \nu \nu)
    = 3 \frac{M_{Z'}^{} g_{X}^2}{24 \pi} q_{f_L^{}}^2~,
\end{align} 
where $q_{f_L^{}}^{}$ is the $U(1)_X$ charge of the SM lepton doublet, $Z^\prime$ gauge boson can decay into a pair of heavy Majorana neutrinos and the partial decay width is given by
\begin{align}
\label{eq:width-NN}
    \Gamma(Z' \to N_R^\alpha N_R^\alpha)
    = \frac{g_{X}^2 M_{Z'}^{}}{24 \pi} x_\nu^2 \left( 1 - \frac{m_{N_i}^2}{M_{Z'}^2} \right)^{\frac{3}{2}}~,
\end{align}
where $x_{\nu}^{}$ is the general $U(1)$ charge of the RHNs and $m_{N_i}$ is its mass. The general $U(1)$ charge of the RHNs could be obtained from Tab.~\ref{tab1}.
\subsection{Heavy neutrino interactions}
\label{HNL}
The neutrino mass in the model is generated by the seesaw mechanism after $U(1)_X$ and electroweak symmetries are broken. Hence the neutrino mass matrix in this scenario can be written as 
\begin{eqnarray}
m_{\nu} \ = \ \begin{pmatrix}
0&&m_{D}\\
m_{D}^{T}&&m_{N}
\end{pmatrix}.  
\label{typeInu}
\end{eqnarray}
where we consider $M_N$ as a diagonal matrix without the loss of generality.  Diagonalizing this mass matrix we obtain the light neutrino mass eigenvalue as
\bea
m_\nu \simeq -m_D m_N^{-1} m_D^T.
\eea
As a result we write down light neutrino flavor eigenstate $(\nu_\alpha)$ in terms of light $(\nu_i)$ and heavy $(N_i)$ mass eigenstates
\bea 
\nu_\alpha \simeq  U_{\alpha i}\nu_i  + V_{\alpha i} N_i,  
\eea 
where $\alpha$ and $i$ are the generation indices. In this expression $U_{\alpha i}$ is a $3 \times 3$ light neutrino mixing matrix which can be expressed as $(1-\frac{\epsilon}{2})U_{\rm PMNS}$ with $\epsilon= V^\ast V^T$, a non-unitary parameter. We define 
\bea
V_{\alpha i} \simeq m_D m_N^{-1}
\label{eq:mixing}
\eea
as the mixing between light and heavy neutrinos. Through the light-heavy mixing, the SM gauge singlet heavy neutrinos interact with the SM gauge bosons. 
Here $U_{\rm PMNS}$ is a $3\times3$ matrix which diagonalizes the light neutrino mass matrix as 
\bea
U_{\rm PMNS}^T~m_\nu~U_{\rm PMNS} = diag(m_1, m_2, m_3).
\eea 
In presence of $\epsilon$ the mixing matrix $U$ becomes non-unitary. Finally under the influence of light-heavy neutrino mixing the charged-current interactions can be expressed in terms of neutrino mass eigenstates as
\bea 
{\mathcal{L}_{CC} \supset 
 -\frac{g}{\sqrt{2}} W_{\mu}
  \bar{e} \gamma^{\mu} P_L   V_{\alpha i} N_i  + \rm{h.c}.}, 
\label{CC}
\eea
where $e$ represents the three generations of the charged leptons. Similarly, in terms of mass eigenstates, the neutral-current interactions can be written as
\bea 
{\mathcal{L}_{NC} \supset 
 -\frac{g}{2 c_w}  Z_{\mu} 
\left[ 
  \overline{N}_m \gamma^{\mu} P_L  (V^{\dagger} V)_{mi} N_i
+ \left\{ 
  \overline{\nu}_m \gamma^{\mu} P_L (U^{\dagger}V)_{mi}  N_i
  + \rm{h.c.} \right\} 
\right] , }
\label{NC}
\eea
 where $c_w \equiv \cos \theta_w$ with $\theta_w$ being the weak mixing angle.
Finally we observe that heavy neutrinos $(N)$ decay into $\ell W$, $\nu Z$ and $\nu h$ respectively where $h$ is the SM Higgs boson. For sterile neutrinos heavier than $W$, $Z$ and $h$, the decays are on-shell, i.e., 2-body followed by the further decays of the SM bosons. The corresponding partial decay widths are 
\bea
\Gamma(N_i \rightarrow \ell_{\alpha} W)
 &=& \frac{|V_{\alpha i}|^{2}}{16 \pi} 
\frac{ (m_{N_i}^2 - M_W^2)^2 (m_{N_i}^2+2 M_W^2)}{m_{N_i}^3 v_h^2} ,
\nonumber \\
\Gamma(N_i \rightarrow \nu_\alpha Z)
&=& \frac{|V_{\alpha i}|^{2}}{32 \pi} 
\frac{ (m_{N_i}^2 - M_Z^2)^2 (m_{N_i}^2+2 M_Z^2)}{m_{N_i}^3 v_h^2} ,
\nonumber \\
\Gamma(N_i \rightarrow \nu_\alpha h)
 &=& \frac{|V_{\alpha i}|^{2}}{32 \pi}\frac{(m_{N_i}^2-M_h^2)^2}{m_{N_i} v_h^2}.
\label{eq:dwofshell}
\eea 
\section{Resonant Leptogenesis}
\label{secIII}
\begin{figure*}[h]
\begin{center}
\includegraphics[width=85mm]{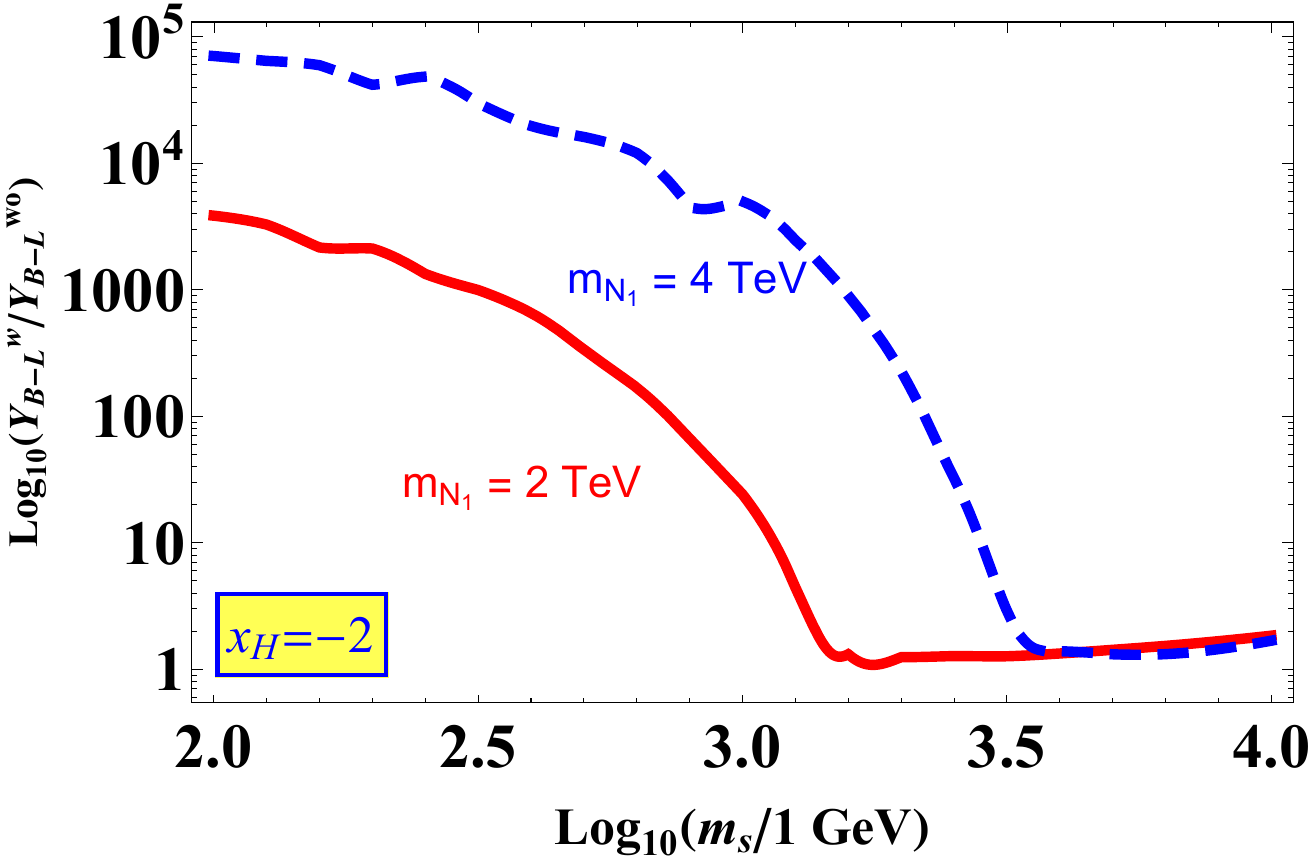} 
\includegraphics[width=85mm]{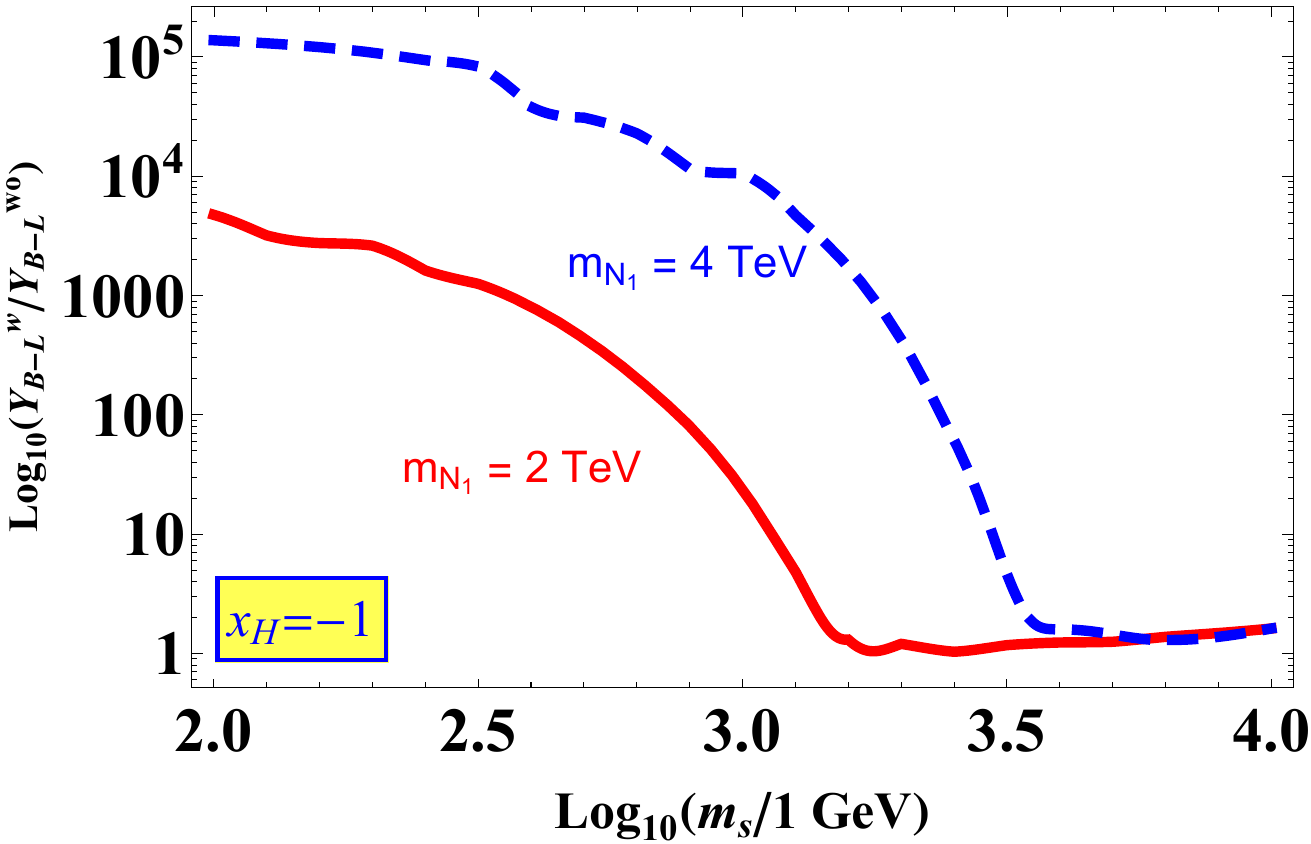}\\
\includegraphics[width=85mm]{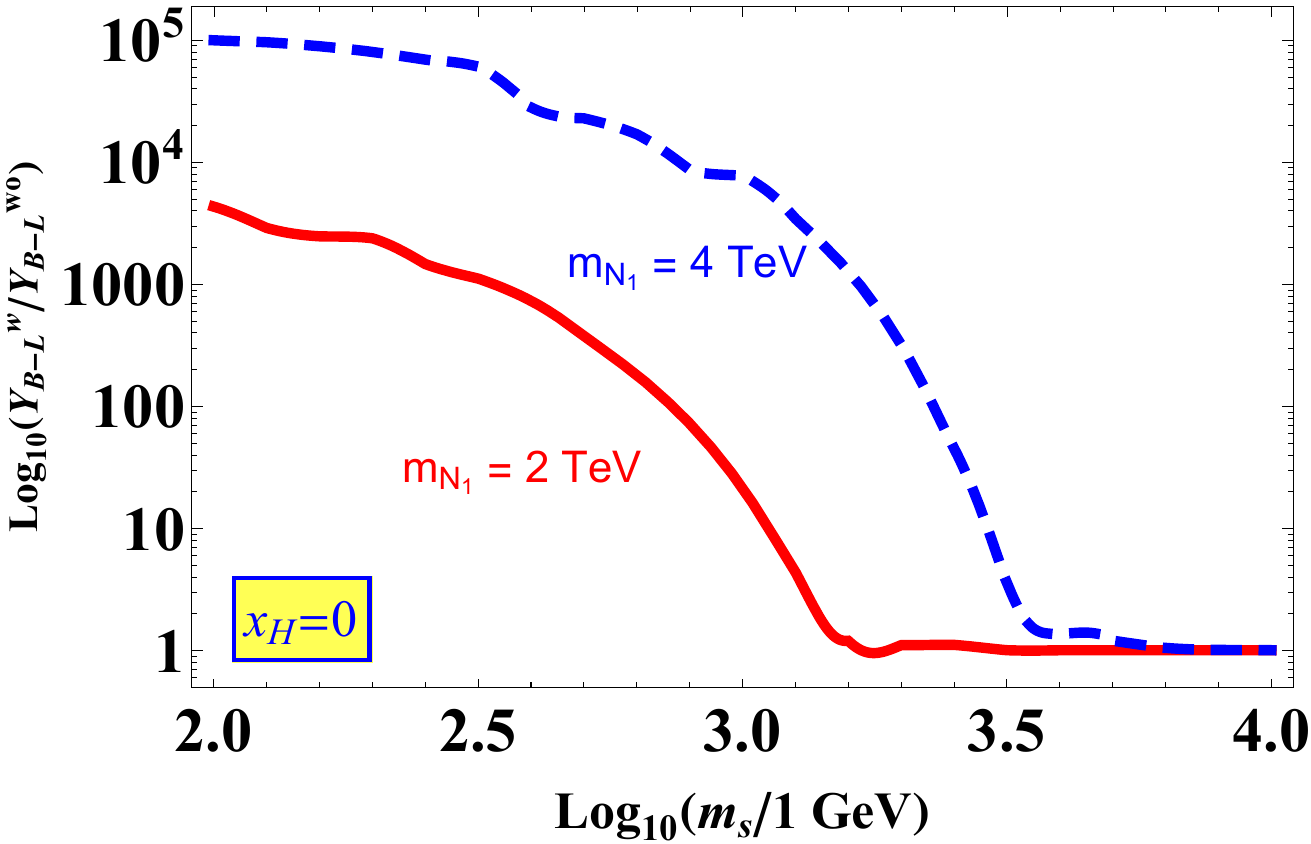} 
\includegraphics[width=85mm]{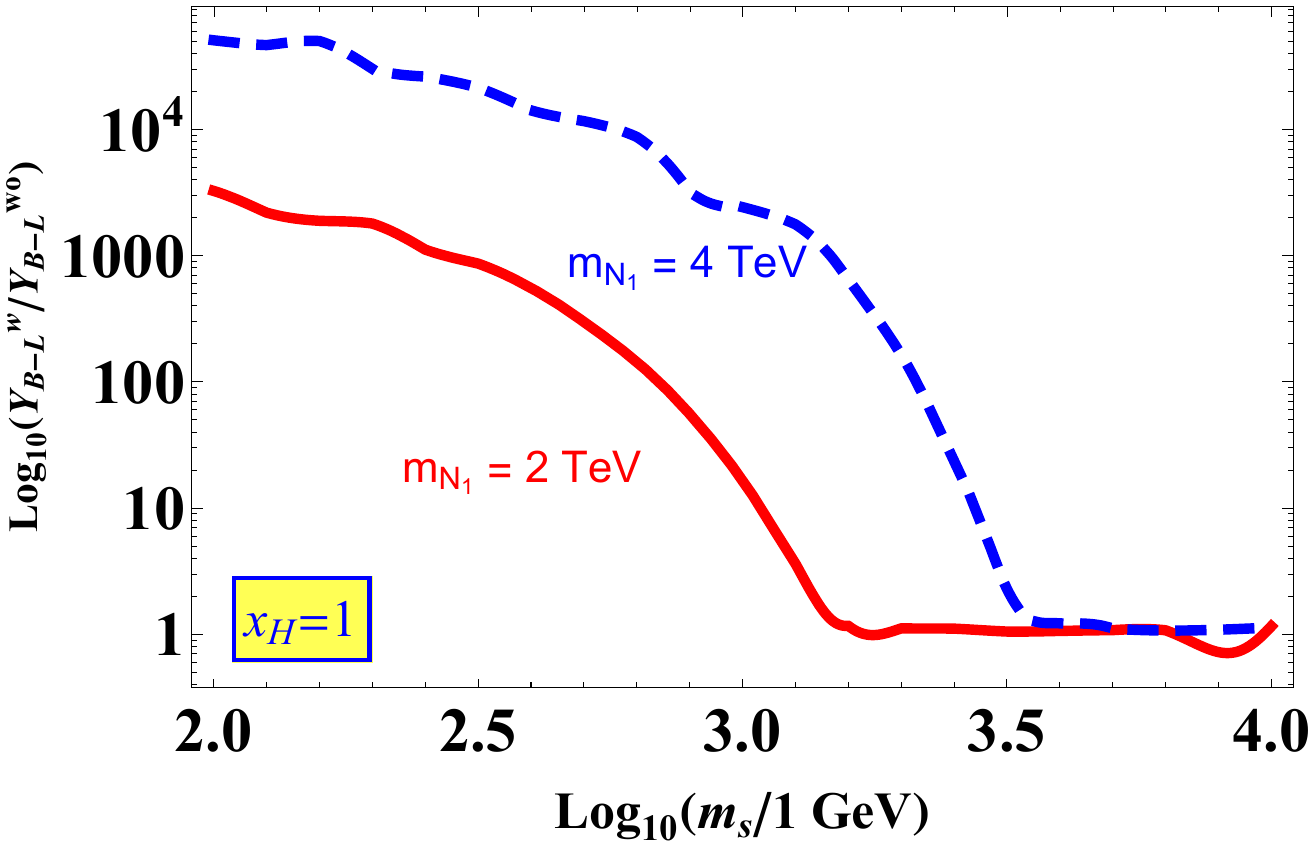}\\
\includegraphics[width=85mm]{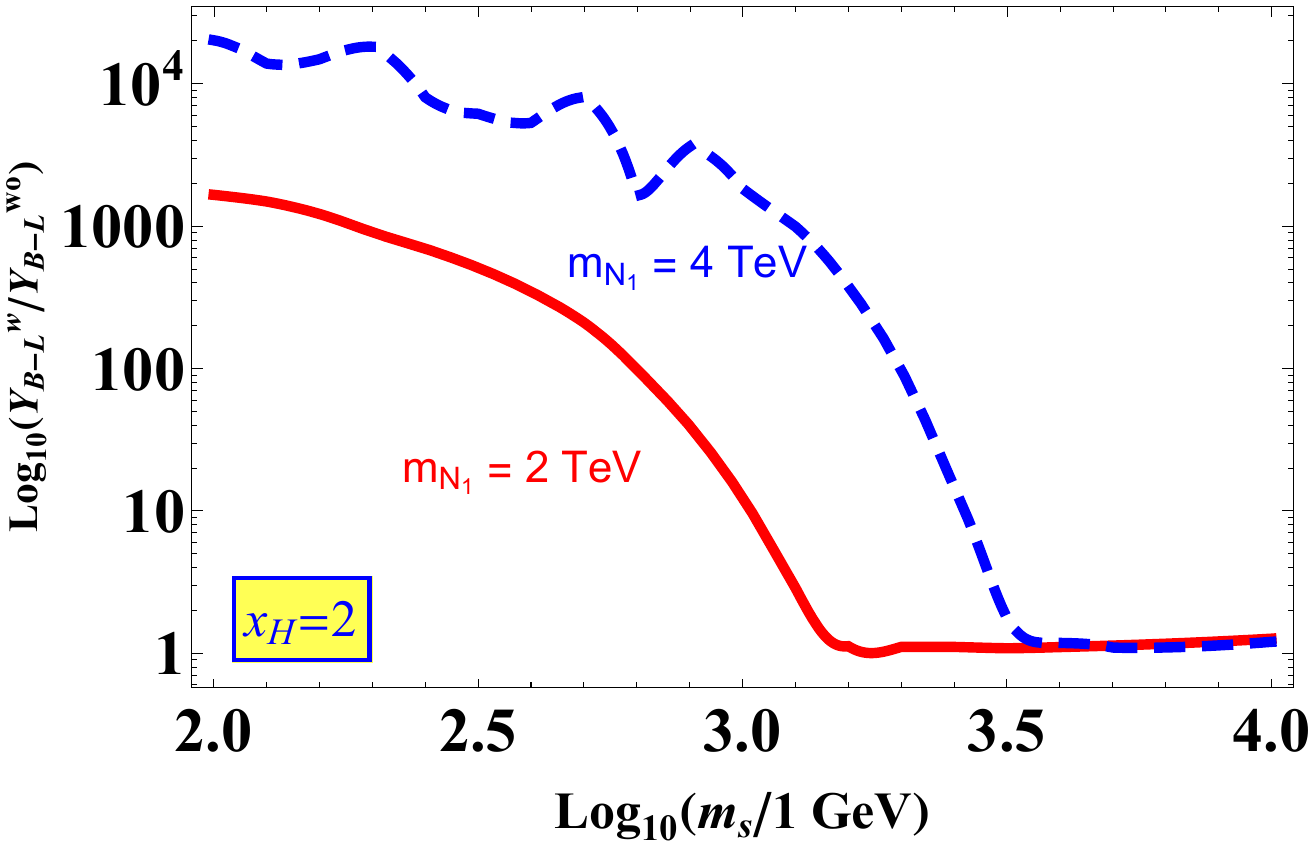} 
 \caption{
 The ratios of the B-L asymmetry with scalar contributions ($Y_{B-L}^w$) and without the contribution ($Y_{B-L}^{wo}$) as functions of the scalar mass.  The red solid line represents $m_{N_1} = 2$ TeV and the blue dashed line represents $m_{N_1} = 4$ TeV. The U(1)$_X$ gauge coupling and $Z^\prime$ boson mass for different $x_H$ fixing $g_X=0.1$ and $M_{Z'} = 6$ TeV.
 }
\label{fig:ratioxh0}
\end{center}
\end{figure*}
The Boltzmann equations that govern the yields of the singlet neutrinos, the B$-$L charge and a scalar are written as
\begin{eqnarray}
\frac{d Y_{N_i}}{dz} &=& - \frac{z}{s H(m_{N_1})}
\left[
\left(\frac{Y_{N_i}}{Y_{N_i}^{eq}}-1\right)
\left(
\gamma_{D_i} + 2 \gamma_{h, s} + 4 \gamma_{h, t}
\right)
+
\left(
\left[\frac{Y_{N_i}}{Y_{N_i}^{eq}}\right]^2-1
\right)
\gamma_{Z'}
\right.
\nonumber\\
&&\left.
+
\left(
\left[\frac{Y_{N_i}}{Y_{N_1}^{eq}}\right]^2
-
\left[\frac{Y_{\Phi}}{Y_{N_1}^{eq}}\right]^2
\right)
\left(
\gamma_{Z', \Phi} + \gamma_{N, \Phi}
\right)
\right], 
\label{boltzN}
\\
\frac{d Y_{\Phi}}{dz} &=& - \frac{z}{s H(m_{N_1})}
\left[
\sum_{i=1}^2
\left(
\left[\frac{Y_{\Phi}}{Y_{N_1}^{eq}}\right]^2
-
\left[\frac{Y_{N_i}}{Y_{N_1}^{eq}}\right]^2
\right)
\left(
\gamma_{Z', \Phi} + \gamma_{N, \Phi}
\right)
+
\left(
\left[\frac{Y_{\Phi}}{Y_{N_1}^{eq}}\right]^2-1
\right)
\gamma_{Z', h}
\right], 
\label{boltzs}
\\
\frac{d Y_{B-L}}{dz} &=& - \frac{z}{s H(m_{N_1})}
\left[
\sum_{i=1}^2\left(
\frac12 \frac{Y_{B-L}}{Y_\ell^{eq}}+
\epsilon_i \left(\frac{Y_{N_i}}{Y_{N_i}^{eq}}-1\right)\right)\gamma_{D_i}
+ \frac{Y_{B-L}}{Y_\ell^{eq}}
\left( 2 \left( \gamma_{N} + \gamma_{N, t} + \gamma_{h, t}\right) +
\sum_{i=1}^2\frac{Y_{N_i}}{Y_{N_i}^{eq}}\gamma_{h, s}
\right)
\right], 
\label{boltzBL}
\end{eqnarray}
\begin{eqnarray}
\epsilon_i &=& -\sum_{j\neq i} \frac{m_{N_i}}{m_{N_j}} \frac{\Gamma_j}{m_{N_j}} 
\left(
\frac{V_j}{2} + S_j
\right)
\frac{{\rm Im} \left[ \left( y_D y_D^\dagger \right)_{ij}^2 \right]}{\left( y_D y_D^\dagger \right)_{ii} \left( y_D y_D^\dagger \right)_{jj}}, 
\\
V_j &=& 2 \frac{m_{N_j}^2}{m_{N_i}^2}\left[ \left(1 + \frac{m_{N_j}^2}{m_{N_i}^2} \right) \log \left(1 + \frac{m_{N_i}^2}{m_{N_j}^2} \right) - 1 \right]
\\
S_j &=& \frac{m_{N_j}^2 \Delta M_{ij}^2}{\left(\Delta M_{ij}^2\right)^2 + m_{N_i}^2 \Gamma_j^2}, 
\ 
\Delta M_{ij}^2 \equiv 
m_{N_j}^2 - m_{N_i}^2, 
\end{eqnarray}
and $\gamma$'s are given by 
\begin{eqnarray}
\gamma_a = \frac{m_N}{64 \pi^4 z}\int ds \hat{\sigma}_a (s) \sqrt{s} K_1 \left(\frac{z \sqrt{s}}{m_N} \right). 
\end{eqnarray}
$\hat{\sigma}_a$'s are reduced cross sections listed in Appendix.~\ref{sec:app1}.

We discuss the contribution of a scalar boson. 
We assume that the scalar mixing is very small and neglect its effects. 
The scalar boson couples to the $Z^\prime$ boson and right-handed neutrinos. 
For the $Z^\prime$ exchange process, the scalar boson is in thermal equilibrium and the decoupling temperature is 
similar to that of the right-handed neutrino. 
Therefore the deviation of the scalar density from thermal equilibrium changes the baryon asymmetry in our universe. 
Fig.\ref{fig:ratioxh0} shows the ratio of the B$-$L asymmetry with a scalar contribution and without a scalar contribution. 
When the scalar boson mass is less than 1 TeV, the scalar contribution is very large. 
We discuss the CP-asymmetry parameter. 
We consider resonant leptogenesis and two neutrino masses are degenerated: 
\begin{eqnarray}
\Delta M_{12}^2 = m_{N_1} \Gamma_2. 
\end{eqnarray}
We assume that the third generation of the right-handed neutrino is a DM candidate and the rank of the neutrino mass matrix is two. 
In the case, the dirac neutrino mass matrix can be written the following using the Casas-Ibarra parametrization \cite{Casas:2001sr}, 
\begin{eqnarray}
m_D = U_{MNS}^\ast 
\left(
\begin{array}{ccc}
\sqrt{m_{1}} & 0 & 0 \\
0 & \sqrt{m_{2}} & 0 \\
0 & 0 & \sqrt{m_{3}}\\
\end{array}
\right)
O
\left(
\begin{array}{ccc}
\sqrt{m_{N_1}} & 0 & 0 \\
0 & \sqrt{m_{N_2}} & 0 \\
0 & 0 & \sqrt{m_{N_3}}\\
\end{array}
\right)
\end{eqnarray}
where 
\begin{eqnarray}
m_1 = 0, m_2 = m_{2NH}, m_3=m_{3NH}, \\
O = 
\left(
\begin{array}{ccc}
0 & 0 & 1 \\
\cos\alpha & \sin\alpha & 0 \\
-\sin\alpha & \cos\alpha & 0 \\
\end{array}
\right), 
\end{eqnarray}
for the NH case and 
\begin{eqnarray}
m_1 = m_{1IH}, m_2 = m_{2IH}, m_3=0, \\
O = 
\left(
\begin{array}{ccc}
\cos\alpha & \sin\alpha & 0 \\
-\sin\alpha & \cos\alpha & 0 \\
0 & 0 & 1 \\
\end{array}
\right), 
\end{eqnarray}
for the IH case.
$\alpha$ is a complex phase. 

In the NH case, the CP-asymmetry parameter is given by observed values and a complex phase $\alpha = \alpha_r + i \alpha_i$:
\begin{eqnarray}
|\epsilon_1|  &=&\left| \frac12 \frac{{\rm Im} \left[ \left( m_D m_D^\dagger \right)_{ij}^2 \right]}{\left( m_D m_D^\dagger \right)_{ii} \left( m_D m_D^\dagger \right)_{jj}}\right|
\\
 &=&
 \left|\frac{ (m_{2NH}^2 - m_{3NH}^2) \sin(2 \alpha_r) \sinh(2 \alpha_i)}{(m_{2NH} - m_{3NH})^2 \cos(
   2 \alpha_r)^2 - (m_{2NH} + m_{3NH})^2 \cosh(2 \alpha_i)^2}\right|
\\
&\leq& \frac{m_{3 NH} - m_{2 NH}}{2(m_{3 NH} + m_{2 NH})}
\label{eq:cpasymnh}
\\
&=& 0.353,  
\end{eqnarray}
where $m_{2NH} =0.00861\ {\rm eV}, m_{3NH} = 0.0502\ {\rm eV}$ \cite{Esteban:2020cvm}. 
The equality of Eq.~(\ref{eq:cpasymnh}) holds when $\alpha = \pm\frac{\pi}{4} + \frac{i}{2} \log(1 + \sqrt{2})$. 
We use the maximum CP-asymmetry parameter in this paper. 

Next, we discuss the CP-asymmetry parameter for the IH case. 
The maximum CP-asymmetry parameter is given by 
\begin{eqnarray}
|\epsilon_1| 
 &=&
 \left|\frac{ (m_{1IH}^2 - m_{2IH}^2) \sin(2 \alpha_r) \sinh(2 \alpha_i)}{(m_{1IH} - m_{2IH})^2 \cos(
   2 \alpha_r)^2 - (m_{1IH} + m_{2IH})^2 \cosh(2 \alpha_i)^2}\right|
\\
&\leq& \frac{m_{2IH} - m_{1IH}}{2(m_{2IH} + m_{1IH})}
= 0.00377,  
\end{eqnarray}
where $m_{1IH} =0.0492\ {\rm eV}, m_{2IH} = 0.0500\ {\rm eV}$ \cite{Esteban:2020cvm}. 
The maximum value is very smaller than NH case and the other contributions are very small, 
therefore, the generated baryon asymmetry becomes very small. 
We don't consider the IH case in this paper.

\section{Results and discussions}
\label{secIV}
\begin{figure*}
\begin{center}
\includegraphics[scale=0.205]{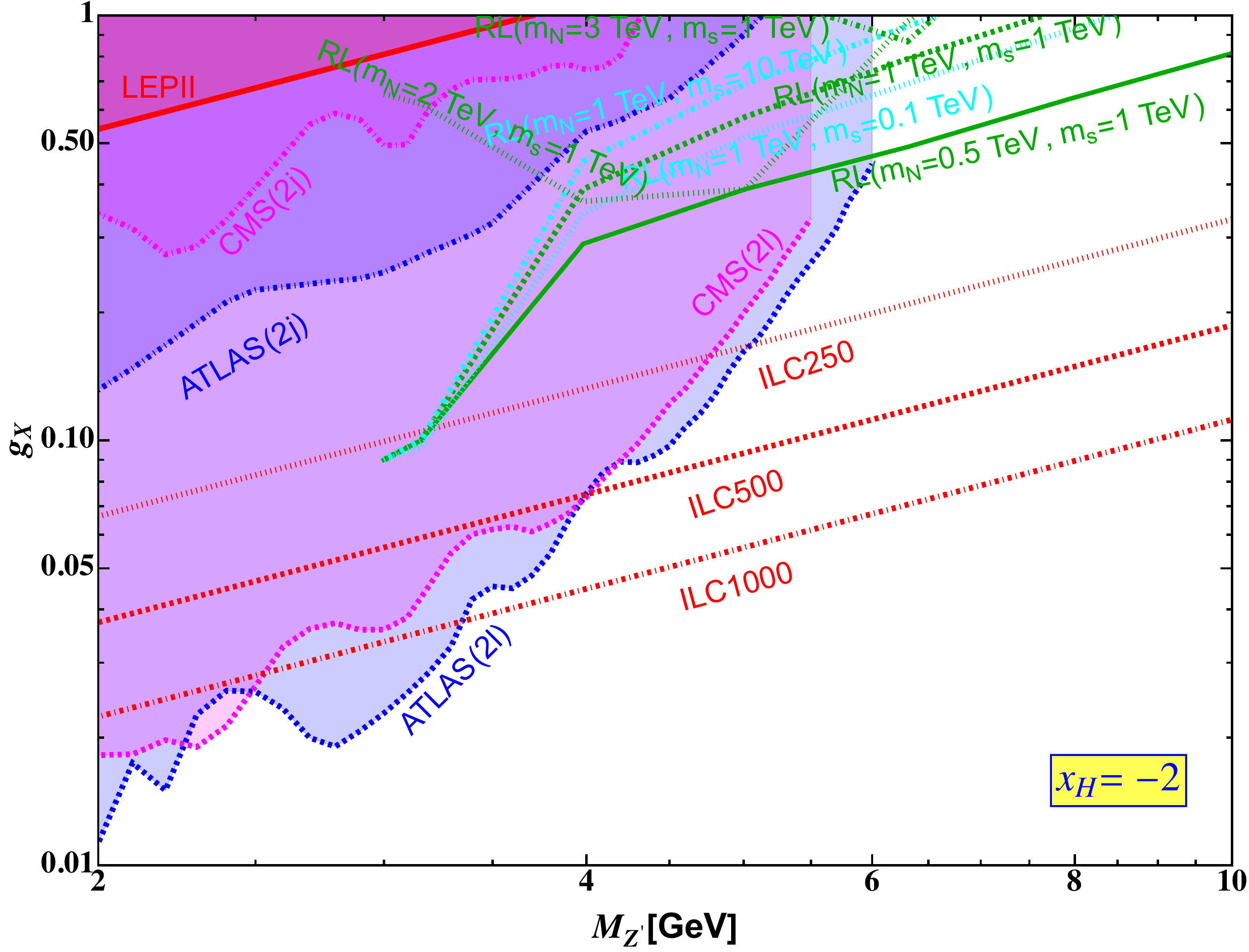}
\includegraphics[scale=0.205]{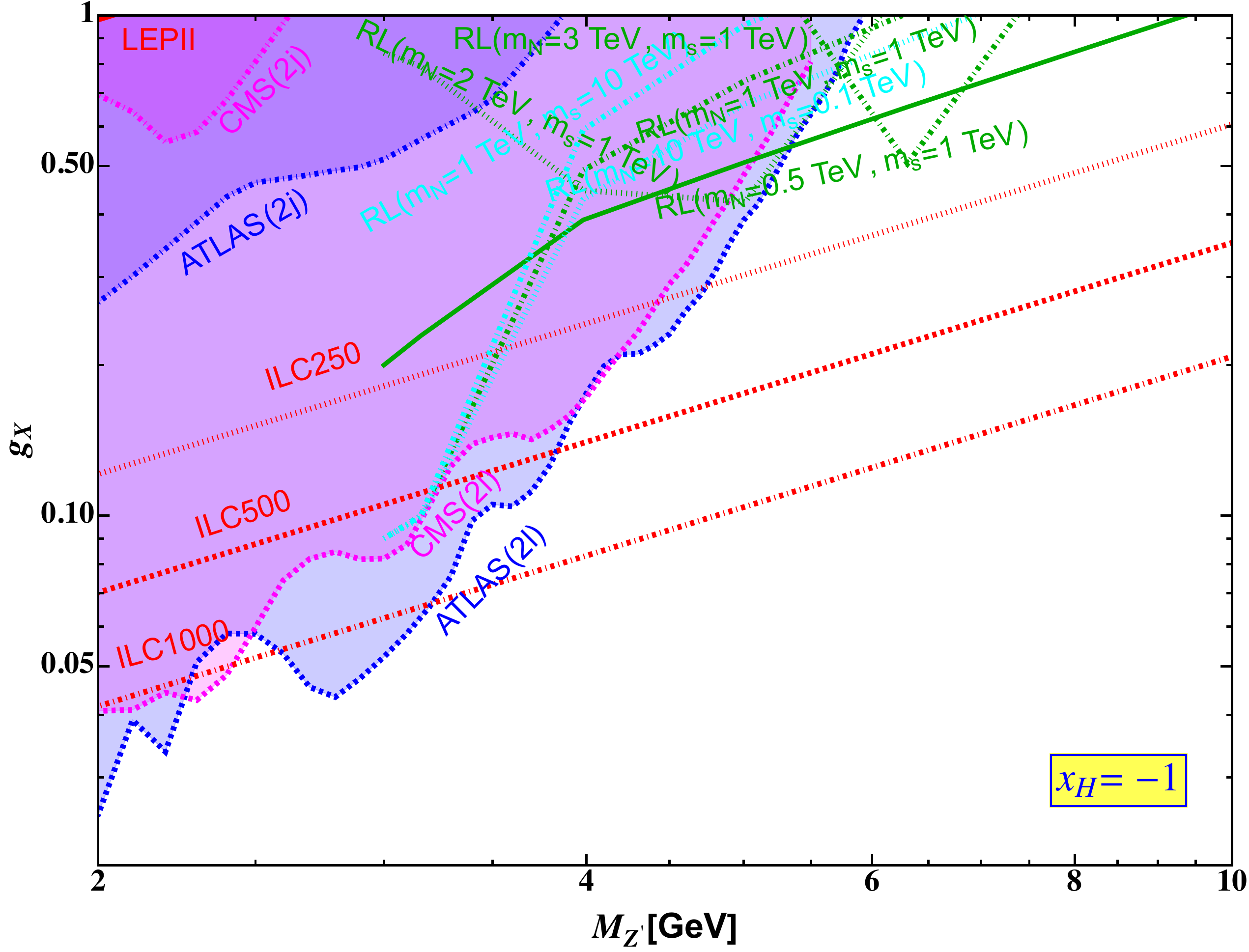}
\includegraphics[scale=0.205]{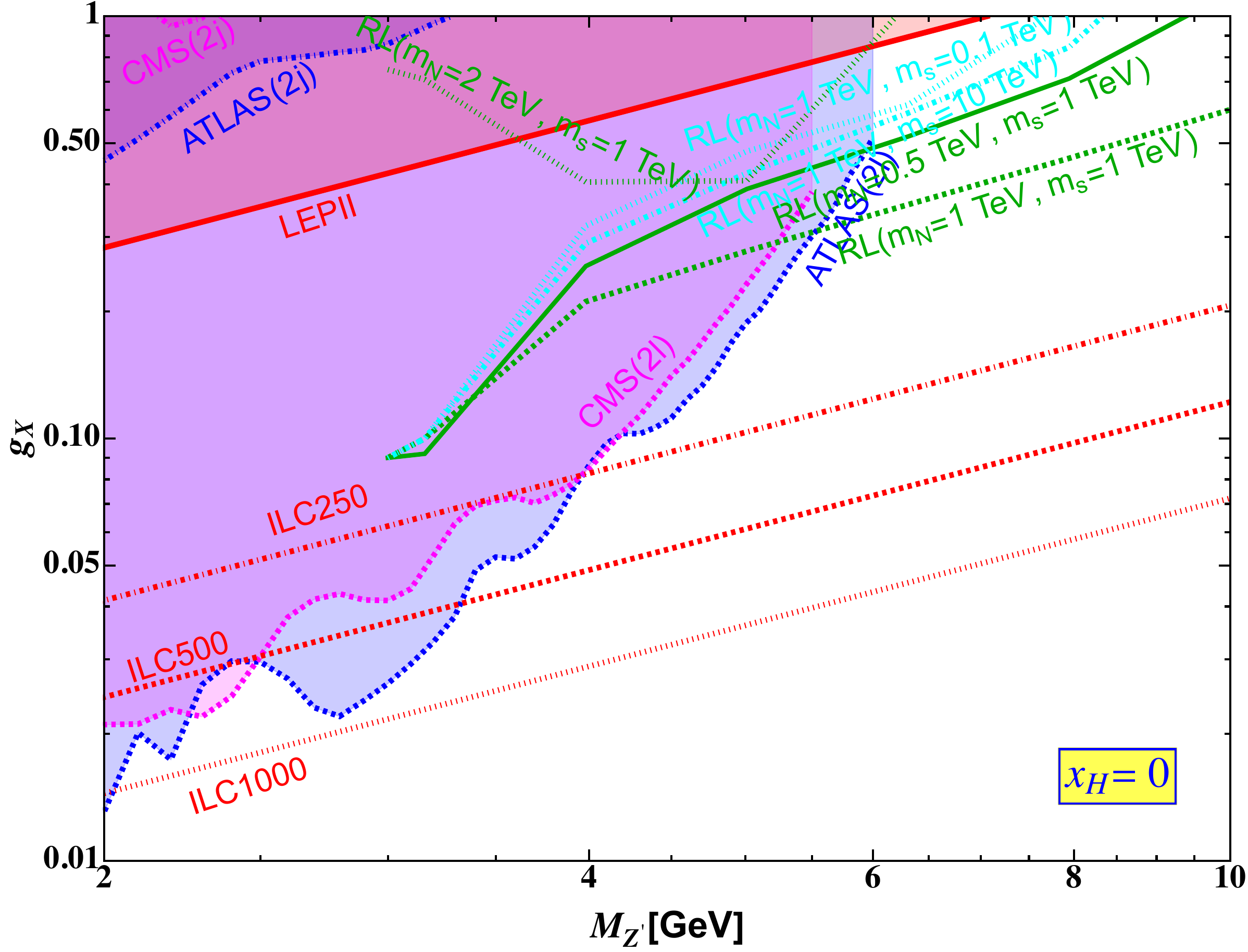}
\includegraphics[scale=0.205]{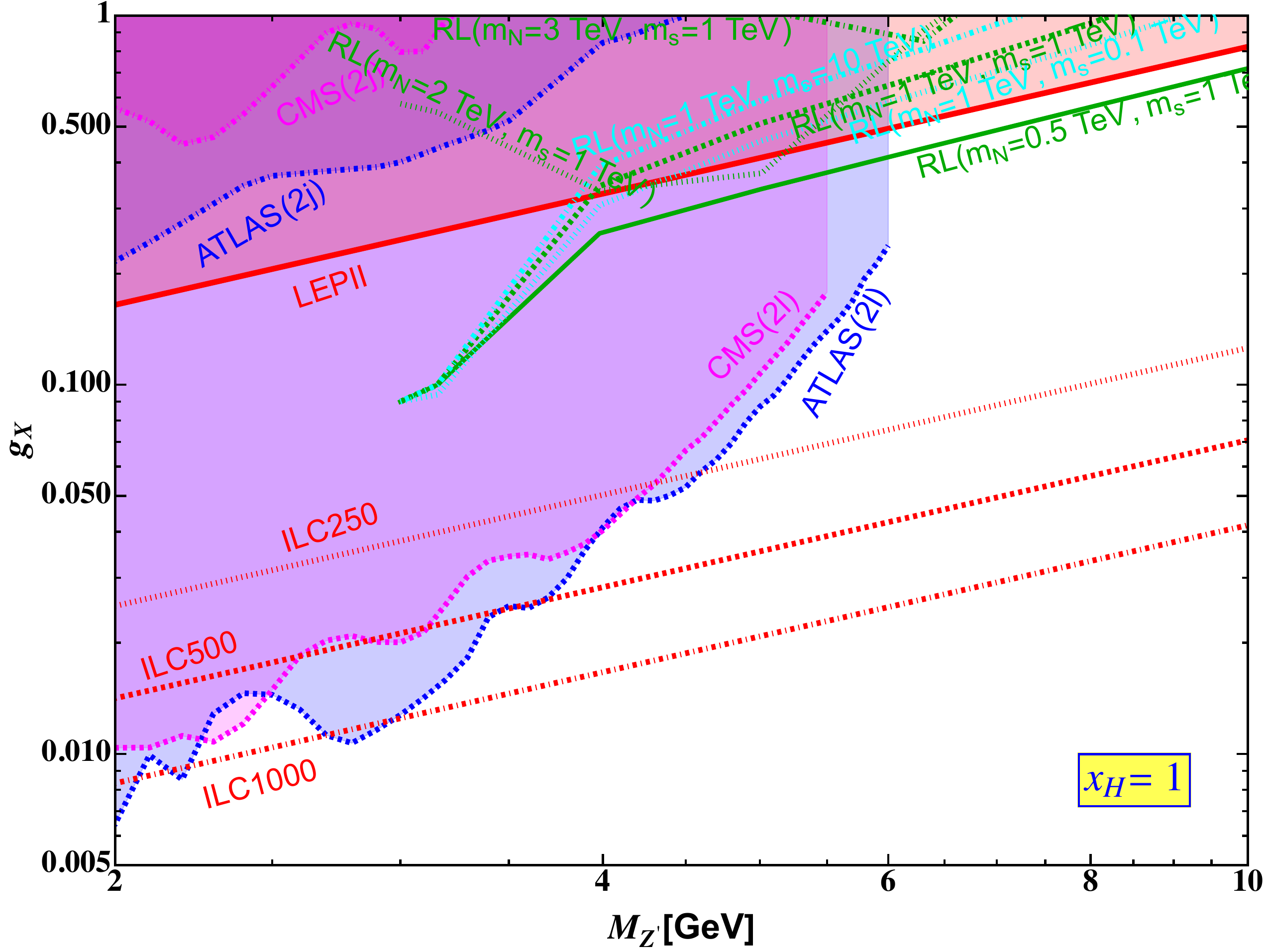}
\includegraphics[scale=0.205]{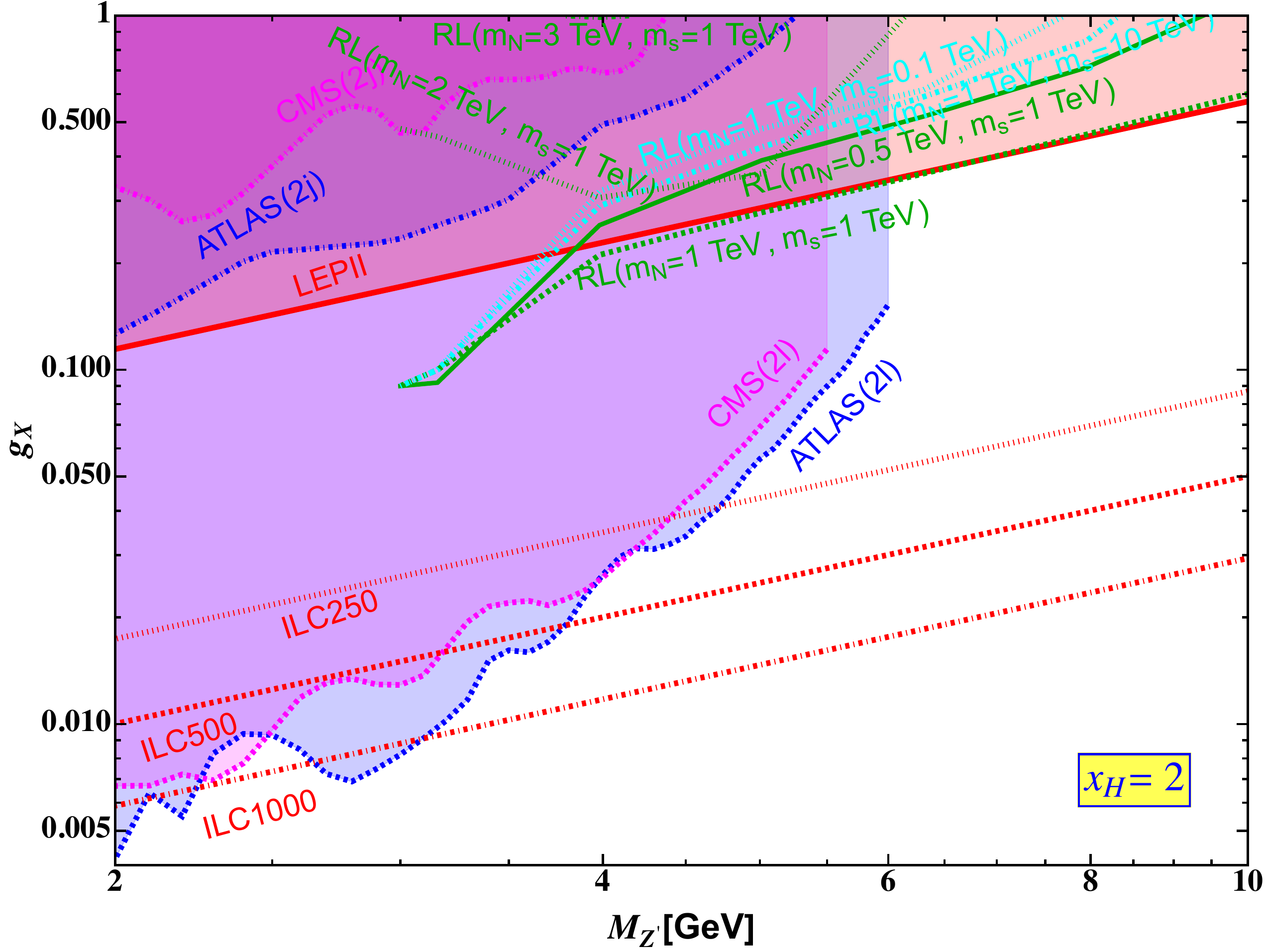}
\caption{Limits from dilepton (dotted), diject (dot-dashed) searches of CMS (magenta), ATLAS (blue), LEP-II (red, solid) and prospective ILC experiments for $\sqrt{s}=250$ GeV (red, dotted), 500 GeV (red, dashed) and 1 TeV (red, dot-dashed). We show the limits from resonant leptogenesis (RL) scenario (green and cyan) for different benchmarks of RHNs $(m_N)$ and scalar $(m_s)$ masses varying $x_H$. We represent RL limes for fixed scalar mass $m_s=1$ TeV with varying RHN mass $m_N=0.5$ TeV (solid green), $m_N=1$ TeV (green dashed), $m_N=2$ TeV (green dotted) and $m_N=3$ TeV (green dot-dashed), respectively. RL limits for fixed RHN mass at $m_N=1$ TeV with different scalar mass $m_s=10$ TeV (cyan dot-dashed) and $m_s=0.1$ TeV (cyan dotted), respectively}
\label{lim1}
\end{center}
\end{figure*}

We numerically solved the Boltzman equations given in Eqs.~(\ref{boltzN}), (\ref{boltzs}) and (\ref{boltzBL}) from $z_{min}=0.001$ to $z_{max}=m_{N_1}/T_{sph}$ with initial conditions: 
\begin{eqnarray}
Y_{N_i}(z_{min}) &=& Y_{N_i}^{eq}(z_{min}), \\
Y_{\Phi}(z_{min}) &=& Y_{\Phi}^{eq}(z_{min}), \\
Y_{B-L}(z_{min}) &=& 0, 
\end{eqnarray}
where $i$ runs 1 and 2, $T_{sph} = 150$ GeV is the sphaleron decoupling temperature \cite{Burnier:2005hp}. 
We fixed the complex phase at $\alpha = \pi/4 + i/2 \log(1+\sqrt{2})$ and the mass of the second right-handed neutrino at $m_{N_2}^2 = m_{N_1}^2 + m_{N_1} \Gamma_2$ as discussed in Sec. \ref{secIII}. 
The B$-$L density is converted to the baryon number density ($Y_B$) through the sphaleron process \cite{Kuzmin:1985mm} by the following formula: 
\begin{eqnarray}
Y_B = \frac{28}{79} Y_{B-L} (z_{max}). 
\end{eqnarray} 
The green and cyan lines in the Fig.~\ref{lim1} represent the observed baryon number density lines, $Y_B= 8.7\times 10^{-11}$, with different Majorana mass and scalar mass. 
The regions above the lines don't create a sufficient baryon asymmetry because we use the maximum CP-asymmetry parameter. Limits obtained from the resonant leptogenesis (RL) scenario are shown in Fig.~\ref{lim1} in darker green and cyan limes for fixed scalar mass $m_s=1$ TeV with varying RHN masses $m_N=0.5$ TeV (solid green), $m_N=1$ TeV (green dashed), $m_N=2$ TeV (green dotted) and $m_N=3$ TeV (green dot-dashed), respectively. We also show the limits for the RL scenario for fixed RHN mass at $m_N=1$ TeV with different scalar masses $m_s=10$ TeV (cyan dot-dashed) and $m_s=0.1$ TeV (cyan dotted), respectively.
Depending on the choice of $x_H$, the constraints obtained for different choices of $m_N$ and $m_s$ vary for a range of the $Z^\prime$ mass $3$ TeV $M_{Z^\prime} \leq 10$ TeV. 

Now we estimate bounds on $g_X$ for different $M_{Z^\prime}$ from the LEP-II searches \cite{Eichten:1983hw,Electroweak:2003ram,ALEPH:2013dgf} for different $x_H$ considering $M_{Z^\prime} \gg \sqrt{s}$ utilizing contact interaction for  $e^-e^+ \to f \bar{f}$ process as 
\bea
{\cal L}_{\rm eff} \ = \ \frac{g_X^2}{(1+\delta_{ef}) (\Lambda_{AB}^{f\pm})^2} \sum_{A,B=L,R}\eta_{AB}(\overline{e} \gamma^\mu P_A e)(\overline{f} \gamma_\mu P_B f) \, ,
\label{eq1}
\eea
where by convention we take $g_X^2/4\pi$ as 1 by where $\delta_{ef}=1\ (0)$ for $f=e$ ($f\neq e$). Here $\eta_{AB}=\pm 1$ or 0, and $\Lambda_{AB}^{f\pm}$ is assumed to be the scale of contact interaction where constructive (destructive)  interference with SM processes $e^+e^-\to f\bar{f}$ \cite{Kroha:1991mn,Carena:2004xs} are represented by plus (minus) sign. The $Z^\prime$ exchange matrix element under $U(1)_X$ scenario can be written as
\bea
\frac{g_X^2}{{M_{Z^\prime}}^2-s} [\overline{e} \gamma^\mu (\tilde{x}_\ell P_L+ \tilde{x}_e P_R) e] [\overline{f} \gamma_\mu (\tilde{x}_{f_L} P_L+ \tilde{x}_{f_R} P_R) f] \, ,
\label{eq2}
\eea
where $\tilde{x}_{f_L}$ and $\tilde{x}_{f_R}$ are the $U(1)_X$ charges of left handed and right handed fermions $(f_L,~f_R)$, respectively from Table~\ref{tab1}. Matching Eqs.~\eqref{eq1} and \eqref{eq2} we find 
\bea
M_{Z^\prime}^2  \ \gtrsim \ \frac{{g_X}^2}{4\pi} |{x_{e_A}} x_{f_B}| (\Lambda_{AB}^{f\pm})^2 \, , 
\label{Lim}
\eea
taking for LEP-II scenario $M_{Z^\prime}^2 \gg s$ where $\sqrt{s}=209$ GeV. Hence we estimate bounds on $M_{Z^\prime}/g_X$ for different values of  $\Lambda_{AB}^{f\pm}$ at $95\%$ for different $x_H$ from \cite{ALEPH:2013dgf} assuming universality in the contact interactions where $AB=LL,\ RR, \ LR, \ RL, \ VV$ and $AA$. The upper limits on $M_{Z^\prime}/g_X$ are $5$ TeV, $2.2$ TeV, $7.0$ TeV $11.1$ TeV and $18$ TeV for $x_H=-2$, $-1$, $0$, $1$ and $2$, respectively which are ruled out by LEP-II data at $\sqrt{s}=209$ GeV. In the same line we estimate prospective limits on $M_{Z^\prime}/g_X$ for different $x_H$ at the ILC with different center of mass energies $\sqrt{s}=250$ GeV, $500$ GeV and $1$ TeV considering limits on $\Lambda_{AB}^{f\pm}$ from \cite{LCCPhysicsWorkingGroup:2019fvj} at $95\%$ C. L. The prospective upper bounds on $M_{Z^\prime}/g_X$ at $\sqrt{s}=250$ GeV for $x_H=-2$, $-1$, $0$, $1$ and $2$ are $31.6$ TeV, $16.3$ TeV, $48.2$ TeV, $79.0$ TeV and $113.7$ TeV, respectively. The prospective upper bounds on $M_{Z^\prime}/g_X$ at $\sqrt{s}=500$ GeV for $x_H=-2$, $-1$, $0$, $1$ and $2$ are estimated as $54.4$ TeV, $26.3$ TeV, $81.6$ TeV, $139.1$ TeV and $199.7$ TeV, respectively. Finally we estimate prospective upper limits on $M_{Z^\prime}/g_X$ for $x_H=-2$, $-1$, $0$, $1$ and $2$ as $88.6$ TeV, $47.7$ TeV, $137.2$ TeV, $238.2$ TeV and $339.2$ TeV, respectively at $\sqrt{s}= 1$ TeV. Hence the estimated lines are shown in Fig.~\ref{lim1} for LEPII (red, solid), ILC250 (red, dotted), ILC500 (red, dashed) and ILC1000 (red, dot-dashed), respectively. 

We calculate limits on $g_X-M_{Z^\prime}$ plane in the $U(1)_X$ scenario for different $x_H$ from the dilepton and dijet searches in ATLAS and CMS \cite{ATLAS:2019erb,CMS:2019tbu} experiments of the LHC. The corresponding cross sections for these scenarios are estimated as $\sigma_{\rm Model}$ involving the $Z^\prime$ contribution for different $x_H$ from $U(1)_X$ model considering a trial value of the general $U(1)_X$ gauge coupling $g_{\rm{Model}}$ at $13$ TeV varying $M_{Z^\prime}$. Comparing these estimated cross sections with the observed cross sections from the LHC $(\sigma_{\rm{Obs.}})$ to estimate limits on the gauge coupling for different $M_{Z^\prime}$ and $x_H$ following 
\bea
g_X \ = \ \sqrt{g_{\rm{Model}}^2 \left(\frac{\sigma_{\rm{Obs.}}}{\sigma_{\rm{Model}}}\right)}.
\label{gp}
\eea
to estimate $95\%$ constraints on the $g_X-M_{Z^\prime}$. To estimate the cross sections of $Z^\prime$ production we used narrow width approximation (NWA) as
\bea
\sigma(pp \to Z^\prime) = 2 \sum_{q, \overline{q}} \int dx \int dy q(x, Q) \overline{q}(y, Q) \hat{\sigma}(\hat{s})
\label{Xsec-1}
\eea
where $q(\bar{q}) (x, Q)$ is the parton distribution functions of the (anti)quark and $\hat{s}= xys$ is the invariant mass squared of the colliding quark at the center of mass energy $\sqrt{s}$ considering up-type and down-type quarks. Due to NWA, the cross section of the colliding quarks to produce $Z^\prime$ boson is 
\bea
\hat{\sigma}=\frac{4\pi^2}{3} \frac{\Gamma(Z^\prime \to q\overline{q})}{M_{Z^\prime}}  \delta(\hat{s}-M_{Z^\prime})
\label{Xsec-2}
\eea
with a factorization scale at $Q=M_{Z^\prime}$ and employing CTEQ6L \cite{Pumplin:2002vw} as the parton distribution function. Partial decay widths of $Z^\prime$, used in this analysis, into a pair of charged fermions are calculated as 
\begin{align}
\label{eq:width-ll}
    \Gamma(Z' \to \bar{f} f)
    &= N_C^{} \frac{M_{Z^\prime}^{} g_{X}^2}{24 \pi} \left[ \left( q_{f_L^{}}^2 + q_{f_R^{}}^2 \right) \left( 1 - \frac{m_f^2}{M_{Z^\prime}^2} \right) + 6 q_{f_L^{}}^{} q_{f_R^{}}^{} \frac{m_f^2}{M_{Z^\prime}^2} \right] \sqrt{1-4 \frac{m_f^2}{M_{Z^\prime}^2}},
\end{align}    
where $m_f$ stands for the mass of the SM fermions, $N_C^{}$ is the color factor which is $1 (3)$ for the SM leptons (quarks) and $q_{f_L(f_R)}$ are the $U(1)_X$ charges of the left (right) handed  charged fermions of the model. That into a pair of light neutrinos $(\nu_L)$ is calculated as
\begin{align}   
\label{eq:width-nunu}
    \Gamma(Z' \to \nu \nu)
    = \frac{M_{Z^\prime}^{} g_{X}^2}{24 \pi} q_{\nu_L^{}}^2~,
\end{align} 
neglecting the tiny mass of the light neutrinos. In this model, the $Z^\prime$ gauge boson can also decay into a pair of heavy Majorana neutrinos when the $Z^\prime$ is heavier than twice the mass of the heavy neutrinos $(m_N)$ which is a free parameter. 
The corresponding partial decay width of $Z^\prime$ into single generation of heavy neutrino pair is given by
\begin{align}
\label{eq:width-NN}
    \Gamma(Z^\prime \to N_R N_R)
    = \frac{M_{Z^\prime}^{} g_{X}^2}{24 \pi} q_{N_R^{}}^2 \left( 1 - 4 \frac{m_N^2}{M_{Z^\prime}^2} \right)^{\frac{3}{2}}~,
\end{align}
with $q_{N_R^{}}^{}$ is the general $U(1)_X$ charge of the heavy neutrinos, and in our case, it is $q_{N_R^{}}^{}=-x_\Phi=-1$. 
If we consider $m_N > M_{Z^\prime}/2$, then $Z^\prime \to N_R N_R$ mode will be kinematically forbidden.  We estimate $\sigma_{\rm{Model}}=$ $\sigma(pp\to Z^\prime) \times$ BR$(Z^\prime \to 2\ell)$ for dilepton and $\sigma(pp\to Z^\prime) \times$ BR$(Z^\prime \to 2j)$ for dijet final states, respectively. Total decay width of $Z^\prime$ used in this analysis has been calculated using the partial decay widths following Eqs.~(\ref{eq:width-ll})- (\ref{eq:width-NN}) to obtain the respective branching ratios. Finally bounds on $g_X-M_{Z^\prime}$ plane are calculated from Eq.~(\ref{gp}) and shown in Fig.~\ref{lim1}. The dilepton bounds provide strong constraints for different $x_H$ compared to the dijet bounds. 

Finally from Fig.~\ref{lim1} we compare the limits obtained from resonant leptogenesis and collider searches for different $x_H$. We find that limits obtained from resonant leptogenesis become stronger than the limits obtained from LEP-II, LHC for $M_{Z^\prime} > 5.8$ TeV for $x_H=-2$ and $-1$. The constraints obtained from LEP-II become stronger than the CMS limits for $x_H \leq 0$, ruling out some parts of the resonant leptogenesis parameters for $x_H=0$ and $1$, respectively. We find that For $x_H=2$, LEP-II limits rule out the limits obtained from the resonant leptogenesis. We find that prospective limits in case of 250 GeV ILC can be comparable with the limits obtained by the resonant leptogenesis scenario. Hence these limits could be probed at 250 GeV ILC for $x_H=-1$. Other prospective limits from ILC at different center of mass energies more than 250 GeV and different $x_H$ could provide stronger constraints than the limits obtained from resonant leptogenesis. 
\section{Conclusions}
\label{secV}
We consider a general $U(1)_X$ scenario where we have three generations of SM-singlet RHNs whichc are charged under general $U(1)_X$ scenario. After cancelling gauge and mixed gauge-gravity anomalies we find that left and right handed charged leptons interact differently with $Z^\prime$. These scenarios affect the generation of CP asymmetry mediated by $Z^\prime$ and scalars while induced by SM and BSM fermions and scalars applying resonant leptogenesis. Reproducing the CP asymmetry applying different benchmark scenarios of the RHN and SM-singlet BSM scalar masses we estimate the bounds on $g_X-M_{Z^\prime}$ plane for different general $U(1)_X$ charges. We estimate limits on general $U(1)_X$ coupling for different $M_{Z^\prime}$ using $M_{Z^\prime} > \sqrt{s}$ for LEP-II and prospective ILC using electron-positron scattering. Comparing with the estimated dilepton and dijet cross sections at the proton-proton collider with the LHC searches we estimate limits on $g_X-M_{Z^\prime}$ plane for different $U(1)_X$ charges. We find that limits obtained from resonant leptogenesis provide stronger bounds compared to those obtained from LHC for $M_{Z^\prime} > 5.8$ TeV. Existing LEP-II limits from electron-positron collisions will provide stronger limits when $x_H \geq 0,$ 1 and partially rule out a part of the parameter regions found using resonant leptogenesis scenario when $M_{Z^\prime} > 5.8$ TeV. Limits obtained from LEP-II for $x_H=2$ provide stronger bounds compared to resonant leptogenesis scenario. We find that for $x_H=-1$ and $-2$ limits obtained from resonant leptogenesis are stronger than any collider limits for $M_{Z^\prime} > 5.8$ TeV. These parameter regions involving TeV scale RHNs, BSM scalar and $Z^\prime$ gauge boson depending on different general $U(1)_X$ charges could be probed at high energy colliders experiments in future. 
\begin{appendix}
\section{Reduced cross sections}
\label{sec:app1}
\begin{widetext}
The reduced cross sections involved in this analysis can be given by $\hat{\sigma}=\sigma \{\frac{8}{s}[(p_f. p_i)^2-m_f^4]\}$ where $\sigma$ is the total cross section in CM frame for different $2 \to 2$ processes participating in leptogenesis, $p_i$ and $p_f$ are momenta of initial and final state particles. We write the reduced cross sections below:
\begin{itemize}
\item[(i)] RHN pair production from different initial states involving SM charged fermions:
\bea
\hat{\sigma}_{Z'}\equiv
\hat{\sigma}(f \bar{f} \leftrightarrow N_i N_i )= \frac{g_X^4 x_\nu^2 s^2}{12 \pi } \Bigg[\frac{(1-\frac{4 m_{N_i}^2}{s})^{\frac{3}{2}}}{(s-{M_{Z^\prime}}^2)^2+ {\Gamma_{Z^\prime}}^2 {M_{Z^\prime}}^2}\Bigg] (C_A^2+ C_V^2)
\label{RHN-1}
\eea
where the vector $(C_V= \frac{q_{f_L}+q_{f_R}}{2})$ and axial-vector $(C_A= \frac{q_{f_L}-q_{f_R}}{2})$ couplings can be derived using the general $U(1)$ charges of the left and right handed fermions from Tab.~\ref{tab1}. The center of mass cross section will be averaged over color factor for quarks $3^2$ in initial states. We calculate the reduced cross sections for the scalar $(h)$ initialed process $h h \to f \bar{f}$ mediated by $Z^\prime$ in the following as
\bea
\hat{\sigma}_{Z', h}\equiv
\hat{\sigma}(h h \leftrightarrow f  \bar{f} )= \frac{g_X^4 x_h^2 s^2 }{12 \pi } \Big[\frac{(1-\frac{4m_h^2}{s})^{\frac{3}{2}}}{{(s-{M_{Z^\prime}}^2)^2+ {\Gamma_{Z^\prime}}^2 {M_{Z^\prime}}^2}}\Big](C_A^2+ C_V^2).
\eea
where $x_h$ and $x_\nu$ are the general $U(1)_X$ charges of the scalars and RHNs. In addition to that, we also calculate the reduced cross section for the scalar initiated process $h h \to h_{\rm SM} h_{\rm SM}$ into a pair of SM Higgs $(h_{\rm SM})$ in the following as 
\bea
\hat{\sigma}(h h \leftrightarrow h_{\rm SM} h_{\rm SM} )= \frac{g_X^4 x_h^2 x_{h_{\rm SM}}^2 s^2}{12 \pi } \frac{(1-\frac{4m_h^2}{s})^{\frac{3}{2}} (1-\frac{4 m_{h_{\rm SM}}^2}{s})^{\frac{3}{2}}}{{(s-{M_{Z^\prime}}^2)^2+ {\Gamma_{Z^\prime}}^2 {M_{Z^\prime}}^2}}.
\eea
where $m_{h_{\rm SM}}$ and $x_{h_{\rm SM}}$ are the mass and general $U(1)$ charge of $h_{\rm SM}$, respectively. 

\item[(ii)] RHN pair production from different generation of RHNs: 
\bea
\hat{\sigma}(N_i N_i \leftrightarrow N_j N_j)=
\frac{g_X^4 x_\nu^4\sqrt{(s-4 m_{N_j}^2)(s-4 m_{N_i}^2)}}{72 \pi s \{(s-M_{Z^\prime}^2)^2+\Gamma_{Z^\prime}^2 M_{Z^\prime}^2\}}
\Bigg[(s-4 m_{N_j}^2)(s-4 m_{N_i}^2)+ 12 \frac{m_{N_i}^2 m_{N_j}^2}{M_{Z^\prime}^4} (s- M_{Z^\prime}^2)^2 \Bigg]
\label{RHN-2}
\eea
where $i\neq j$. RHN pair production form same generation:
\bea
\hat{\sigma}(N_i N_i \leftrightarrow N_i N_i )&=& \frac{ g_x^4 x_\nu^4}{128\pi}\Bigg[\frac{(s-4 m_{N_i}^2)^3}{3 s((s-M_{Z^\prime}^2)^2+M_{Z^\prime}^2 \Gamma_{Z^\prime}^2)}+ \frac{(s-4 m_{N_i}^2)}{s M_{Z^\prime}^4 (s-4 m_{N_i}^2+ M_{Z^\prime}^2)} \Big\{M_{Z^\prime}^2(s-4 m_{N_i}^2)^2+ \nonumber \\ 
&&2 (M_{Z^\prime}^2 -2 m_{N_i}^2)^3+ s(8 m_{N_i}^4+ 3 M_{Z^\prime}^4)-4 m_{N_i}^2 (M_{Z^\prime}^4+ 4 m_{N_i}^4+ M_{Z^\prime}^2 m_{N_i}^2)\Big\}+ \nonumber \\
&&+\Big\{\frac{(3 M_{Z^\prime}^2-4 m_{N_i}^2)(s-4 m_{N_i}^2)^2+ M_{Z^\prime}^4(3s-20 m_{N_i}^2)+ 2 M_{Z^\prime}^2 (M_{Z^\prime}^4+8 m_{N_i}^4)}{s M_{Z^\prime}^2 (s-4 m_{N_i}^2+ 2 M_{Z^\prime}^2)}\Big\}\nonumber \\ 
&& \log\Big(\frac{M_{Z^\prime}^2}{s-4 m_{N_i}^2+ M_{Z^\prime}^2}\Big)\Bigg]~~~~~
\eea
\item[(iii)] RHN pair production from $h$ in $s-$channel:
\bea
\hat{\sigma}_{Z', \Phi}\equiv
\hat{\sigma}(h h \leftrightarrow N_i N_i)= \frac{{g_X^4} x_h^2 x_\nu^2 s^2}{12 \pi} \frac{(1-\frac{4 m_{N_i}^2}{s})^{\frac{3}{2}} (1-\frac{4 {m_h}^2}{s})^{\frac{3}{2}}}{M_{Z^\prime}^2 \Gamma_{Z^\prime}^2+(s-M_{Z^\prime}^2)^2}.
\eea
Reduced cross section for $h$ pair production from RHN in $t-$channel and $u-$channel processes:
\bea
\hat{\sigma}_{N, \Phi}\equiv
\hat{\sigma}(N_i N_i \leftrightarrow h h)= \frac{Y_N^4}{8 \pi} \Bigg(1-\frac{4 m_{N_i}^2}{s}\Bigg)\Bigg[\frac{s-4m_{N_i}^2}{2 m_{N_i}^2}+\Bigg(1-\frac{4 m_{N_i}^2}{s}\Bigg)^{\frac{3}{2}} \log \Bigg(\frac{s-\sqrt{s(s-4 m_{N_i}^2)}}{s+\sqrt{s(s-4 m_{N_i}^2)}}\Bigg)\Bigg]~~~~~
\eea
\item[(iv)] Scalar mediated $\ell N^i \to t q$ process in $s-$channel:
\bea
\hat{\sigma}_{h, s}\equiv
\hat{\sigma}(N^i \ell \leftrightarrow t q)=
\frac{3 \pi \alpha^2 m_t^2}{M_W^4 \sin\theta_W^4} (m_D^\dagger m_D)_{ij} \Bigg(1-\frac{m_{N_i}^2}{s}\Bigg)^2
\eea
Scalar mediated $ N^i t \to \ell q$ process in $t-$channel:
\bea
\hat{\sigma}_{h, t}\equiv
\hat{\sigma}(N^j t \to \ell q)= 
\frac{3 \pi \alpha^2 m_t^2}{M_W^4 \sin\theta_W^4} (m_D^\dagger m_D)_{ij} \Bigg[1-\frac{m_{N_i}^2}{s}+\frac{m_{N_i}^2}{s}\log\Bigg(1+\frac{s-m_{N_i}^2}{m_h^2}\Bigg)\Bigg]
\eea
where $\alpha=\frac{e^2}{4 \pi}$.
\item[(v)] RHN mediated $\ell h \to \ell h$ process in $s-$channel and $t-$ channel:
\bea
\hat{\sigma}_{N}\equiv
\hat{\sigma}(\ell h \leftrightarrow \ell h)&=& 
\frac{2\alpha^2 \pi}{ \sin^4\theta_W M_W^4 s} \Bigg\{\sum_{i=1}^{2} m_{N_{i}}^2 (m_D^\dagger m_D)_{ii}^2
\Bigg[\frac{s}{m_{N_i}^2}+ \frac{2 s}{\mathcal{D}_i}+\frac{s^2}{2 \mathcal{D}_i^2}
-\Bigg(1+2\frac{s+m_{N_i}^2}{\mathcal{D}_i}\Bigg)\log\Bigg(1+\frac{s}{m_{N_i}^2}\Bigg)\Bigg]
\nonumber \\
&&
+2 m_{N_1}m_{N_2}\mathcal{R}e\Bigg[(m_D^\dagger m_D)_{12}^2\Bigg] \Bigg[\frac{s}{\mathcal{D}_1}+\frac{s}{\mathcal{D}_2}+\frac{s^2}{2 \mathcal{D}_1\mathcal{D}_2}-
\frac{(s+m_{N_1}^2)(s+m_{N_1}^2-2 m_{N_2}^2)}{(m_{N_1}^2-m_{N_2}^2)\mathcal{D}_2} \log\Bigg(1+\frac{s}{m_{N_1}^2}\Bigg)
\nonumber \\
&&
-
\frac{(s+m_{N_2}^2)(s+m_{N_2}^2-2 m_{N_1}^2)}{(m_{N_2}^2-m_{N_1}^2)\mathcal{D}_1} \log\Bigg(1+\frac{s}{m_{N_2}^2}\Bigg)
\Bigg]
\Bigg\}
\eea
where $\mathcal{D}_i= \frac{(s-m_{N_i}^2)^2+ m_{N_i}^2  \Gamma_{i}^2}{(s-m_{N_i}^2)}$ is the off-shell part of the propagator and it is a dimensionless quantity. 
\item[(vi)] RHN mediated $\ell \ell \leftrightarrow h h $ process in $t-$ channel: 
\bea
\hat{\sigma}_{N, t}\equiv 
\hat{\sigma}(\ell \ell \leftrightarrow h h)&=& \frac{2 \pi \alpha^2}{M_{W}^4 \sin\theta_W^4} \Bigg\{\sum_{i=1}^{2} (m_D^\dagger m_D)_{ii}^2 \Bigg[ \frac{s}{2 (s+ m_{N_i}^2)}+\frac{m_{N_i}^2}{s+ 2 m_{N_i}^2} \log\Bigg(1+\frac{s}{m_{N_i}^2}\Bigg)\Bigg]\nonumber \\
&+&\mathcal{R}e\Bigg[(m_D^\dagger m_D)_{12}^2\Bigg] \frac{m_{N_1} m_{N_2}}{(m_{N_1}^2- m_{N_2}^2)(s+ m_{N_1}^2+ m_{N_2}^2)}\Bigg[(s+2 m_{N_1}^2) \log\Bigg(1+\frac{s}{m_{N_2}^2}\Bigg) \nonumber \\
&-& (s+ 2 m_{N_2}^2) \log\Bigg(1+\frac{s}{m_{N_1}^2}\Bigg) 
\Bigg]
\Bigg\}.
\eea
\end{itemize}
\end{widetext}
\end{appendix}
\bibliographystyle{utphys}
\bibliography{ref}
\end{document}